\documentstyle[12pt]{article}
\newcommand{\be}{\begin{equation}}
\newcommand{\ee}{\end{equation}}
\newcommand{\ba}{\begin{eqnarray}}
\newcommand{\ea}{\end{eqnarray}}

\begin{document}
\begin{center}
{\bf NEW TWO-DIMENSIONAL INTEGRABLE QUANTUM MODELS FROM SUSY
INTERTWINING
}\\
\vspace{0.5cm} {\large \bf M.V. Ioffe$^{1,2,}$\footnote{E-mail:
m.ioffe@pobox.spbu.ru}, J. Negro$^{2,}$\footnote{E-mail:
jnegro@fta.uva.es}, L.M. Nieto$^{2,}$\footnote{E-mail:
luismi@metodos.fam.cie.uva.es},
D.N. Nishnianidze$^{1,3,}$\footnote{E-mail: qutaisi@hotmail.com}}\\
\vspace{0.2cm} $^1$ Department of Theoretical Physics, Sankt-Petersburg State University,\\
198504 Sankt-Petersburg, Russia\\
$^2$ Departamento de F\'{\i}sica Te\'orica, At\'omica y \'Optica,
Universidad de Valladolid,
47071 Valladolid, Spain\\
$^3$ Kutaisi State University, 4600 Kutaisi, Georgia
\end{center}
\vspace{0.2cm} \hspace*{0.5in}
\hspace*{0.5in}
\begin{minipage}{5.0in}
{\small Supersymmetrical intertwining relations of second order in
the derivatives are investigated for the case of supercharges with
deformed hyperbolic metric $g_{ik}=diag(1,-a^2)$. Several classes
of particular solutions of these relations are found. The
corresponding Hamiltonians do not allow the conventional
separation of variables, but they commute with symmetry operators
of fourth order in momenta. For some of these
models the specific SUSY procedure of separation of variables is applied.  \\

\vspace*{0.1cm} PACS numbers: 03.65.-w, 03.65.Fd, 11.30.Pb }
\end{minipage}
\vspace*{0.2cm}

\section{Introduction}
\vspace*{0.1cm} \hspace*{3ex} Two-dimensional Supersymmetric
Quantum Mechanics \cite{review} with supercharges of second order
in derivatives was shown to be instrumental in building new
integrable two-dimensional quantum and classical systems which do
not allow the standard separation of variables \cite{david} -
\cite{exact}. All such system have, by construction, symmetry
operators of fourth order in momenta which, as a rule, are not
reducible to lower order. Therefore, these models {\it do not
allow} the conventional separation of variables (for definitions
see \cite{miller}).

The possible intertwining operator - supercharges - can be easily
classified into four inequivalent classes depending on their
metrics $g_{ik}$  \cite{david1}. The intertwining relations of
second order can be represented in the form of a system of six
nonlinear partial differential equations involving two partner
potentials plus six functions defining the supercharge. It does
not seem to be possible to solve this system in a general form.
Till now, only some particular cases of constant metric were
investigated \cite{david} - \cite{exact}: elliptic (Laplacian)
$g_{ik}=\delta_{ik}$, Lorentz $g_{ik}=diag(1,-1),$ and degenerate
$g_{ik}=diag(1,0).$ The elliptic case leads to models with
separation of variables and second order symmetry operators, but
the other gave rise to a list of {\it completely integrable}
models (with fourth order symmetry operators) which are not
amenable to separation of variables. Some of these models were
found previously by other methods \cite{perelomov}, but some of
them are new. We notice that although, in general, there is no
direct relation between the integrability (with symmetry operators
of order higher than two in momenta) and the solvability of the
system, it was shown that some models with this kind of
integrability allow partial \cite{new} or even complete
\cite{exact} solvability by means of SUSY methods.

Only one class of intertwining supercharges {\it with constant
metric} in the supercharges was not yet investigated: the case of
a diagonal matrix $g_{ik}=diag(1,-a^2),\quad a\neq 0,\pm 1.$ Just
the study of this kind of intertwining relation is the objective
of the present paper, the corresponding metric being naturally
called a `deformed hyperbolic' one.

The paper is organized as follows. In Section 2 the solution of
the intertwining relations with deformed hyperbolic metrics is
performed, in such a way that the unknown functions satisfy unique
nonlinear functional-differential equation. Since the general
solution of this equation is very difficult to find, some suitable
ans\"atze are investigated in Section 3, and particular solutions
of the intertwining relations are obtained explicitly. In Section
4 the special case when variables are separated only in one of the
partner Hamiltonians is considered. A new SUSY-algorithm of
separation of variables for the second partner potential is
proposed for this particular class of models. In Section 5 we
consider the solution of the intertwining relations with deformed
hyperbolic metric when one of the partner is chosen to be the
isotropic harmonic oscillator. It is shown that its superpartner
necessarily allows separation of variables too. Some final
conclusions put an end to this paper.

\section{SUSY intertwining relations for supercharges with deformed
hyperbolic metric}
\vspace*{0.1cm} \hspace*{3ex} Let us consider
the SUSY intertwining relations
\begin{equation}H_1Q^+ =
Q^+H_2 \quad\quad Q^-H_1 = H_2Q^- \label{intertw}
\end{equation}
between two two-dimensional partner Hamiltonians of Schr\"odinger
type
\begin{equation}
H_{1,2} = -\triangle + V_{1,2}(\vec x) \qquad \triangle \equiv
\partial_1^2 + \partial_2^2 \qquad
\partial_i \equiv \partial/\partial x_i  \qquad \vec x=(x_1,x_2)\label{defh}
\end{equation}
with second order supercharges of the form
\begin{equation}
Q^+ = g_{ik}(\vec x)\partial_i\partial_k +  C_i(\vec x)
\partial_i + B(\vec x) \quad\quad Q^-=(Q^+)^{\dagger}. \label{ourq}
\end{equation}
These intertwining relations realize the isospectrality (up to
zero modes of $Q^{\pm}$) of the superpartners $H_1, \, H_2$ and
the connection between their wave functions with the same values
of energy:
\begin{equation}
\Psi^{(1)}_n(x) = Q^+ \Psi^{(2)}_n(x) \quad\quad \Psi^{(2)}_n(x) =
Q^- \Psi^{(1)}_n(x) \quad\quad n=0,1,2,... \label{psi}
\end{equation}
Equations (\ref{intertw}) are equivalent \cite{david} to the
following system of six nonlinear partial differential equations
\ba
&\partial_iC_k + \partial_kC_i-(V_1-V_2)g_{ik}=0\label{1}\\
  &\triangle C_i + 2\partial_iB +2g_{ik}\partial_kV_2 - (V_1-V_2)C_i=0 \label{2}\\
  &\triangle B + g_{ik}\partial_i\partial_kV_2 + C_i\partial_iV_2 -(V_1-V_2)B=0 \label{3}
\ea where, for simplicity, the explicit dependence of the
functions on the variables has been eliminated.

For the particular case $g_{ik}=diag(1,-a^2), \,\,a\neq 0,\pm 1,$
considered in this paper, the three equations in (\ref{1}) take
the form: \ba
  &2\partial_1C_1=v \quad\quad 2\partial_2C_2=-a^2v \label{1.2}\\
  &\partial_1C_2 + \partial_2C_1=0 \label{1.3}
\ea where the notation $v(\vec x)\equiv V_1(\vec x)-V_2(\vec x)$
has been introduced. From Eq.(\ref{1.2}) one can easily express
the three functions $C_{1,2}$ and $ v$ in terms of a unique
arbitrary function $C(\vec x):$
\begin{equation}
  C_1(\vec x)= -\frac{1}{a^2}\partial_2C(\vec x) \quad\quad C_2(\vec x)= \partial_1C(\vec
  x)  \quad\quad v(\vec x)=-\frac{2}{a^2}\partial_1\partial_2C(\vec x).
  \label{C}
\end{equation}

Then, due to (\ref{1.3}), the arbitrary function $C(\vec x)$ must
satisfy the second order wave equation:
\begin{equation}
\biggl(\partial^2_1-\frac{1}{a^2}\partial^2_2\biggr)C(\vec x)=0.
\label{CC}
\end{equation}
The new variables {\bf $x_{\pm}\equiv x_1\pm ax_2$} are obviously
suitable to write the general solution of (\ref{CC}) in terms of
two arbitrary functions $C_{\pm}(x_{\pm})$ as follows
\begin{equation}
C(\vec x)= \int C_+(x_+)dx_+ +  \int C_-(x_-)dx_- . \label{CFF}
\end{equation}
Therefore, from (\ref{C}) we get
\begin{equation}\label{CCv}
  C_1(\vec x)=-\frac{1}{a}(C_+(x_+)-C_-(x_-)) \quad C_2(\vec
  x)=C_+(x_+)+C_-(x_-) \quad v(\vec
  x)=-\frac{2}{a}(C^{\prime}_+(x_+)-C^{\prime}_-(x_-))
\end{equation}
where the prime means derivative of the corresponding function
with respect to its argument (we will also use below the notation
$\partial_{\pm}\equiv \partial / \partial x_{\pm}$).

The two equations in (\ref{2}) can be rewritten now in terms of
the functions $C_+, C_-$ and their derivatives as follows : \ba
  -\frac{1+a^2}{a}(C_+^{\prime\prime}-C_-^{\prime\prime}) + 2(\partial_++\partial_-)(B+V_2)-
  \frac{2}{a^2}(C_+^{\prime}-C_-^{\prime})(C_+-C_-)&=&0 \label{22}\\
    \frac{1+a^2}{a}(C_+^{\prime\prime}+C_-^{\prime\prime})+2(\partial_+-\partial_-)(B -
    a^2V_2)+ \frac{2}{a^2}(C_+^{\prime}-C_-^{\prime})(C_++C_-)&=&0. \label{222}
\ea Simple linear combinations of (\ref{22}) and (\ref{222}) lead
to the system:
\ba
  &a\partial_+(2B+(1-a^2)V_2)=-\frac{2}{a}(C_+^{\prime}-C_-^{\prime})C_--(1+a^2)C_-^{\prime\prime}-
  a(1+a^2)\partial_-V_2 \label{5}\\
&a\partial_-(2B+(1-a^2)V_2)=\frac{2}{a}(C_+^{\prime}-C_-^{\prime})C_++(1+a^2)C_+^{\prime\prime}-
  a(1+a^2)\partial_+V_2.\label{6}
\ea The consistency condition for these two equations is:
\begin{equation}
\partial_1\partial_2 \Biggl[a^2(1+a^2)V_2 -(C_+^2+C_-^2)-
a(1+a^2)(C_+^{\prime}-C_-^{\prime})\Biggr]=0. \label{consist}
\end{equation}
Therefore, the partner Hamiltonians can be expressed in terms of
the four arbitrary functions $C_+(x_+), C_-(x_-), F_1(x_1)$ and
$F_2(x_2)$: \ba
  H_{1,2} &=&-\Delta
  +V_{1,2}(\vec x) \quad \Delta = \partial_1^2+\partial_2^2=
  (1+a^2)(\partial_+^2+\partial_-^2)+2(1-a^2)\partial_+\partial_-\label{hh}\\
  V_{1,2}(\vec
  x)&=&\mp\frac{1}{a}(C_+^{\prime}(x_+)-C_-^{\prime}(x_-))+\frac{1}{a^2(1+a^2)}(C_+^2(x_+)+C_-^2(x_-))+\nonumber\\
 &&+ F_1(x_1)+F_2(x_2). \label{7}
\ea

From (\ref{5}), (\ref{6}) and (\ref{7}), one obtains the
expression for $B:$ \ba
  B(\vec
  x)&=&-\frac{1}{a^2}C_+(x_+)C_-(x_-)-\frac{1-a^2}{2a}(C^{\prime}_+(x_+)-C^{\prime}_-(x_-))-\nonumber\\
  &&-\frac{1-a^2}{2a^2(1+a^2)}(C_+^2(x_+)+C_-^2(x_-))-
  (F_1(x_1)-a^2F_2(x_2)).\label{8}
\ea

Thus, all these algebraic manipulations transformed the initial
problem of solving the system of differential equations (\ref{1})
- (\ref{3}) to expressions (\ref{CCv}), (\ref{7}) and (\ref{8}) in
terms of arbitrary functions $C_+(x_+), C_-(x_-), F_1(x_1),
F_2(x_2)$, which are restricted by the only remaining equation
(\ref{3}): \ba &\partial_+\{b C''_+(x_+)+2C_+(x_+)[C_+^2(x_+)
+C_-^2(x_-)+2F(\vec{x})]\}=\nonumber\\
&=\partial_-\{b C''_-(x_-)+2C_-(x_-)[C_+^2(x_+)
+C_-^2(x_-)+2F(\vec{x})]\} \label{star} \ea where function $F$ and
constant $b$ were defined as
\begin{equation}F(\vec
x)\equiv\frac{a^2(1+a^2)}{1-a^2}(F_1(x_1)-a^2F_2(x_2)) \qquad
b\equiv -a^2(1+a^2)^2 .\label{K}
\end{equation}
The connection with previous works on the Lorentz metric case
\cite{david} - \cite{exact} can be established very easily.
Indeed, Eq. (\ref{star}) should be multiplied by $(1-a^2)$, and
the limit $a^2\to 1$ should be taken. Then, the result given in
Eq. (13) of \cite{new} is straightforwardly obtained.

 Since Eq.(\ref{star}) is the functional-differential
equations for the functions $C_{\pm}, F_{1,2}$, which depend on
their own arguments, there are no chances to solve this equation
in the most general form. Nevertheless, in  next section some
particular solutions of (\ref{star}) will be found by suitably
chosen ans\"atze. In the following we are mainly interested in the
models where at least one of the partner Hamiltonians is not
amenable to separation of variables.

\section{Particular solutions of intertwining relations}
\vspace*{0.1cm} \hspace*{3ex} In order to obtain particular
solutions of the functional-differential equation (\ref{star}),
some extra hypothesis must be done. We will consider the following
three cases.

\subsection{Case in which $C_-(x_-)=0$}

A great simplification is obtained if we choose one of the
functions $C_{\pm}$ to be zero. Without loss of generality, let us
consider $C_-(x_-)=0.$ Then, one has to solve the second order
equation for the functions $C_+(x_+), F(\vec x):$
\begin{equation}\label{first}
  b C''_+(x_+)+2C_+^3(x_+)+4C_+(x_+)F(\vec{x})=U_-(x_-)
\end{equation}
where  $U_-$ is an arbitrary function. Different possibilities
appear here:

\begin{enumerate}
\item {$U_-(x_-)=Const$}

From (\ref{first}), this choice gives immediately  that $F(\vec
x)$ actually depends only on $x_+$. Therefore, from (\ref{K}),
$F_{1,2}(x_{1,2})$ are at most linear functions of the
corresponding arguments, and $F(\vec x)=c_+x_+ +c_0$. This form of
$F(\vec x)$ allows separation of variables for both Hamiltonians
(\ref{hh}): to separate variables one has to rewrite (\ref{hh}) in
terms of $z_+=x_+=x_1+ax_2$ and $z_-=ax_1-x_2$. In the variables
$z_{\pm}$ the kinetic part of the Hamiltonian, in contrast to
variables $x_{\pm}$ (see (\ref{hh})), does not contain mixed
terms, and the separation of variables for a linear function $F$
becomes evident.

\item {$U_-(x_-)\neq Const$}

In this case, from Eq. (\ref{first}), one can express $F(\vec x)$
as
\begin{equation}
  4F(\vec x)=U_+(x_+)U_-(x_-)-L_+(x_+) \nonumber
\end{equation}
where the new functions are defined as:
\begin{equation}\label{UL}
  U_+(x_+)\equiv C_+^{-1}(x_+) \quad L_+(x_+)\equiv
  \frac{bC_+^{\prime\prime}(x_+)+2C_+^3(x_+)}{C_+(x_+)}= \frac{2[1+b(U_+^{\prime})^2]}{U_+^2}
  -\frac{bU_+^{\prime\prime}}{U_+}.
\end{equation}

The special form (\ref{K}) of the function $F(\vec x)$ means that
\ba
  &&(\partial_+^2-\partial_-^2)F(\vec{x})=
  U_+''(x_+)U_-(x_-)-U_+(x_+)U_-''(x_-)-L_+''(x_+)=0 \label{UUL}\\
&&\frac{U_+''(x_+)}{U_+(x_+)}=\frac{U_-'''(x_-)}{U_-'(x_-)}\equiv\eta^2\nonumber
\ea where $\eta$ is an arbitrary {\it real or purely imaginary}
constant.

For $\eta =0,$ the functions $U_+(x_+), \,U_-(x_-)$ are
polynomials of first and second order, respectively. After their
substitution into (\ref{UUL}), and comparing with (\ref{UL}), one
can check that both partner Hamiltonians allow separation, either
in the variables $x_{\pm}$ or in the variables $z_{\pm}.$

For $\eta\neq 0$, the generic form of the functions $U_{\pm}$ is:
\ba U_+(x_+)&=&\sigma_+\exp(\eta x_+)+\delta_+\exp(-\eta x_+)
\label{U1}\\
U_-(x_-)&=&\sigma_-\exp(\eta x_-)+\delta_-\exp(-\eta x_-)+\delta
\label{U2} \ea where the constant coefficients
$\sigma_{\pm},\,\delta_{\pm}$ must assure the real chracter of
$U_{\pm}(x_{\pm}).$ Again, comparing (\ref{UUL}) and (\ref{UL}),
after a simple calculation one can obtain that
$L_+=b\eta^2\quad\delta =0;\quad b\sigma_+\delta_+\eta^2=1/4,$ and
\ba F_1(x_1)&=&\frac{1-a^2}{4a^2(1+a^2)}\biggl
(\sigma_+\sigma_-\exp(+2\eta x_1) +\delta_+\delta_-\exp(-2\eta
x_1)\biggr ) +k_1 \label{K+}\\
F_2(x_2)&=&-\frac{1-a^2}{4a^4(1+a^2)}\biggl
(\sigma_+\delta_-\exp(+2a\eta x_2) +\delta_+\sigma_-\exp(-2a\eta
x_2)\biggr ) +k_2 \label{K-} \ea where the constants $k_i$ are
such that $k_1-a^2k_2=(1-a^4)\eta^2/4.$ The expressions (\ref{7})
for the potentials $V_{1,2}$ include also the function $C_+(x_+)$,
which has to be found from (\ref{UL}), i.e.
\begin{equation}
  bC_+''(x_+)+2C_+^3(x_+)-b\eta^2C_+(x_+)=0.\nonumber
\end{equation}

This equation can be integrated once:
\begin{equation}\label{FF}
  b(C^{\prime}_+(x_+))^2=-(C_+^2(x_+)-b\eta^2/2)^2+C \qquad C=const .
\end{equation}
For the arbitrary value of real constant $C$ the function
$C_+(x_+)$ can be expressed in terms of elliptic functions (see
\cite{ODE}), and the corresponding potentials in (\ref{7}) will be
written in terms of such $C_+$ and $F$ from (\ref{K+})-
(\ref{K-}). Both partner Hamiltonians do not allow separation of
variables.

For the specific value $C=0,$ Eq. (\ref{FF}) becomes a couple of
Ricatti equations \cite{ince} (let us remind that by definition
$b=-a^2(1+a^2)^2<0$):
\begin{equation}\label{ricatti}
  \sqrt{-b}C^{\prime}_+(x_+)=\pm(C_+^2(x_+)-b\eta^2/2).
\end{equation}
The solutions are well known:
\begin{equation}
C_+(x_+)=\mp\frac{i\eta\sqrt{-b}}{\sqrt{2}} \frac{\nu\exp(i\eta
x_+/\sqrt{2})-\tilde\nu \exp(-i\eta x_+/\sqrt{2})}{\nu\exp(i\eta
x_+/\sqrt{2})+\tilde\nu \exp(-i\eta x_+/\sqrt{2})} \label{Fnu}
\end{equation}
where the constants $\nu ,\, \tilde\nu$ have to keep $C_+(x_+)$
real.

Choosing the lower sign in (\ref{Fnu}) one obtains from (\ref{7}),
(\ref{K+}) and (\ref{K-}) the following expressions for the
partner potentials (up to a common constant term): \ba
   V_{2}(\vec x)&=& F_1(x_1)+F_2(x_2)-(1+a^2)\eta^2/2 \label{ttg1}
   \\
  &=& \frac{1-a^2}{4a^4(1+a^2)}
   \Biggl[a^2(\sigma_+\sigma_-e^{2\eta x_1} +\delta_+\delta_-e^{-2\eta
x_1})- ( \sigma_+\delta_-e^{2a\eta x_2}
+\delta_+\sigma_-e^{-2a\eta x_2} )\Biggr]  \nonumber
\\
  V_{1}(\vec x)&=& V_2(\vec x)-\frac{2}{a}C^{\prime}_+(x_+)=V_2(\vec
  x)+\frac{2}{a^2(1+a^2)}C^2_+(x_+)+\eta^2.\label{ttg0}
\ea Choosing the other sign in (\ref{Fnu}) is equivalent to the
interchange $V_1\leftrightarrow V_2.$ The constants in potential
$V_1$ can be tuned so that it grows at $|\vec x|\to\infty$, and
the corresponding Schr\"odinger equation will allow separation in
terms of the variables $x_1,x_2.$ The solutions of both
one-dimensional equations can be represented in terms of Mathieu
functions \cite{ODE}. Therefore, the Schr\"odinger equation with
potential $V_1(\vec x)$ in (\ref{ttg0}), which does not allow
separation of variables, is isospectral (up to zero modes of
supercharges) to the Schr\"odinger equation with potential $V_2$
in (\ref{ttg1}). Since this last potential allows obviously
separation of variables, one obtains a specific SUSY-separation of
variables for potential (\ref{ttg0}). For further developments of
this model, see Section 4 below.
\end{enumerate}

\subsection{Two Ricatti equations}

The second ansatz which will give some solutions of the
intertwining relations, i.e. solutions of (\ref{star}), is:
\begin{equation}\label{a2}
C^{\prime}_{\pm}(x_{\pm})=cC_{\pm}^2(x_{\pm})+d
\end{equation}
with arbitrary non-zero real constants $c,d.$ The general
solutions of these Ricatti equations are:
\begin{equation}\label{Fff}
  C_{\pm}=d\cdot \frac{f_{\pm}}{f'_{\pm}} \qquad
f_{\pm}=\sigma_{\pm}\exp(\gamma  x_{\pm})+
\delta_{\pm}\exp(-\gamma  x_{\pm})\qquad \gamma \equiv\sqrt{-cd}
\end{equation}
where we can have $cd>0$ or $cd<0$.

Substituting relations (\ref{a2}) into (\ref{star}), one can check
that this last equation takes a simple form if $ca(1+a^2)=\pm 1,$
indeed :
\begin{equation}\label{FK}
  \partial_+\Biggl(C_+(x_+)\biggl(F(\vec x)-d/c\biggr)\Biggr)=\partial_-\Biggl(C_-(x_-)\biggl(F(\vec
  x)-d/c\biggr)\Biggr)
\end{equation}
with $F(\vec x)$ defined in (\ref{K}). This partial differential
equation for $F$ can be solved in general form (see
\cite{classical}) by introducing the new variables
$t_{\pm}\equiv\int dx_{\pm}/C_{\pm}(x_{\pm}):$
\begin{equation}\label{FFK}
  \partial_{t_+}\bigl(C_+C_-(F-d/c)\bigr) =
  \partial_{t_-}\bigl(C_+C_-(F-d/c)\bigr).
\end{equation}
Therefore, its general solution is expressed in terms of an
arbitrary function $M$ of the combination $(t_++t_-).$ Hence, we
have:
\begin{equation}
\label{KM}
  F(\vec x)-d/c=M\biggl(\int\frac{dx_+}{C_+}+\int\frac{dx_-}{C_-}\biggr)/C_+(x_+)C_-(x_-)
\end{equation}
where both functions $C_{\pm}(x_{\pm})$ are given by (\ref{Fff}).
Then, Eq. (\ref{KM}) reads:
\begin{equation}
  F(\vec x)-d/c=\frac{1}{d^2}M\biggl(\frac{1}{d}\ln(f_+f_-)\biggr )f'_+f'_-/f_+f_-\equiv
  U(f_+f_-)f'_+f'_-. \nonumber
\end{equation}
Now one has to remind the special dependence of $F(\vec x)$ on
$\vec x$ given in (\ref{K}), i.e. $\partial_1\partial_2F(\vec
x)=0$, which means that $U^{\prime\prime}(f_+f_-)=0.$ Finally, we
obtain the following solutions for $F_{1,2}:$ \ba
  F_1(x_1)&=&k_1\biggl(\sigma_+\sigma_-e^{2\gamma x_1}+\delta_+\delta_-e^{-2\gamma
  x_1}\biggr)
+k_2\biggl(\sigma^2_+\sigma^2_-e^{4\gamma  x_1}+
\delta^2_+\delta^2_-e^{-4\gamma  x_1}\biggr)+c_1 \label{k+}
\nonumber\\
F_2(x_2)&=&\frac{k_1}{a^2}\biggl(\sigma_+\delta_-e^{2a\gamma x_2}+
\sigma_-\delta_+e^{-2a\gamma x_2}\biggr)+
\frac{k_2}{a^2}\biggl(\sigma^2_+\delta^2_-e^{4a\gamma x_2}+
\sigma^2_-\delta^2_+e^{-4a\gamma x_2}\biggr)+c_2 \label{k-}
\nonumber \ea where $c_1-a^2c_2=cd(1-a^4).$ If we choose
$c=+1/a(1+a^2)$, the partner potentials can be written in an
explicit form \ba
V_1&=&-\frac{8d\sigma_-\delta_-}{a(\sigma_-e^{\gamma x_-}-
\delta_-e^{-\gamma x_-})^2}+ \label{29}\\&&+
k_1\biggl(\sigma_+\sigma_-e^{2\gamma x_1}+
\delta_+\delta_-e^{-2\gamma x_1}\biggr)+
k_2\biggl(\sigma^2_+\sigma^2_-e^{4\gamma
x_1}+\delta^2_+\delta^2_-e^{-4\gamma x_1}\biggr)+ \nonumber\\&&+
\frac{k_1}{a^2}\biggl(\sigma_+\delta_-e^{2a\gamma x_2}+
\sigma_-\delta_+e^{-2a\gamma x_2}\biggr)+
\frac{k_2}{a^2}\biggl(\sigma^2_+\delta^2_-e^{4a\gamma x_2}+
\sigma^2_-\delta^2_+e^{-4a\gamma x_2}\biggr) \nonumber \ea \ba
V_2&=&-\frac{8d\sigma_+\delta_+}{a(\sigma_+e^{\gamma x_+}-
\delta_+e^{-\gamma x_+})^2}+ \label{299}\\&&+
k_1\biggl(\sigma_+\sigma_-e^{2\gamma x_1}+
\delta_+\delta_-e^{-2\gamma x_1)}\biggr)+
k_2\biggl(\sigma^2_+\sigma^2_-e^{4\gamma
x_1}+\delta^2_+\delta^2_-e^{-4\gamma
x_1}\biggr)+\nonumber\\&&+\frac{k_1}{a^2}\biggl(\sigma_+\delta_-e^{2a\gamma
x_2}+ \sigma_-\delta_+e^{-2a\gamma x_2}\biggr)+
\frac{k_2}{a^2}\biggl(\sigma^2_+\delta^2_-e^{4a\gamma x_2}+
\sigma^2_-\delta^2_+e^{-4a\gamma x_2}\biggr) \nonumber \ea up to a
common additive constant. The alternative choice $c=-1/a(1+a^2)$
corresponds to the interchange $V_1\leftrightarrow V_2.$

We can observe that by choosing one of the following constants
$\{\delta_+, \sigma_+, \delta_-, \sigma_-\}$ to be zero, we will
obtain that only one of the partner Hamiltonians $H_1,\,H_2$
admits separation of variables. For example, $\sigma_+=0$ allows
to separate variables in $H_2$ (but not in $H_1$); the two
one-dimensional potentials, that appear after separation of
variables, are exactly solvable Morse potentials, and their
eigenfunctions are known analytically. The separation of variables
in $H_2$ gives a chance to investigate a whole variety of bound
states for the partner Hamiltonian $H_1$ too. In this sense one
can speak of a specific kind of "SUSY-separation of variables" in
$H_1$ (for other types of SUSY-separation of variables see also
\cite{new,exact} and Section 4 below).

\subsection{Case in which $C^{\prime}_{\pm}(x_{\pm})=c C_{\pm}^2(x_{\pm})$}

 This ansatz is a limit case of the previous one, (\ref{a2})  with
$d=0.$ The corresponding solutions are
\begin{equation}
  C_{\pm}(x_{\pm})=-1/cx_{\pm}. \nonumber
\end{equation}
A calculation similar to the one carried out in the previous
Subsection give us: \ba
  F(\vec x)&=&(x_+x_-) M(x_+^2+x_-^2) \qquad M^{\prime\prime}=0
  \label{00}\\
  F_1(x_1)&=&a_1x_1^2+b_1x_1^4 \qquad F_2(x_2)=a_1x_2^2+b_1a^2x_2^4
  \label{000}\\
  V_{1,2}(\vec x)&=&\frac{2(1+a^2)}{(x_{\pm})^2}
  +a_1(x_1^2+x_2^2)+b_1(x_1^4+x_2^4). \label{0000}
\ea The two isospectral Hamiltonians with potentials (\ref{0000})
do not admit separation of variables\footnote{Due to the
coefficients of the attractive singular terms in (\ref{0000}),
these Hamiltonians are symmetric operators, but they have no
self-adjoint extensions (see \cite{reed}).}.

\section{SUSY-separation of variables}
\vspace*{0.1cm} \hspace*{3ex} As we noticed in the examples worked
out in the previous Section, there are some models for which one
of the partner Hamiltonians allows separation of variables, but
the other does not. This means that the solution of
two-dimensional Schr\"odinger equation for the separable
Hamiltonian (for example, $H_2$) is reduced to the solution of two
separate one-dimensional Schr\"odinger problems. Its wave
functions are expressed as a bilinear combination of these
one-dimensional wave functions with arbitrary constant
coefficients. We can use the SUSY intertwining relations
(\ref{psi}) to obtain the wave functions $\Psi_n^{(1)}$ of $H_1$,
one can use $\Psi_n^{(1)}$ from the now known wave functions
$\Psi_n^{(2)},$ acting with the supercharge $Q^+.$ But in contrast
to the one-dimensional situation, some additional eigenstates of
$H_1$ may exist: if they are annihilated by $Q^-$, there are no
partners in the spectrum of $H_2$. Therefore, our task now is to
find all the normalizable wave functions of $H_1$, which are
simultaneously the zero modes of $Q^-.$

\subsection{The algorithm}

To resolve this problem in the case of separation of variables in
$H_2,$ we will use the following trick: though the variables are
separated neither in $H_1,$ nor in $Q^-,$ we will consider a
linear combination:
\begin{equation}\label{linear}
  Z\equiv \alpha H_1+\beta Q^- \qquad \alpha , \beta = const.
\end{equation}

Let us suppose now the existence of some constants $\alpha , \beta
, $ such that the variables in the operator $Z$ are separated by
some similarity transformation, and its normalizable
eigenfunctions could be found. Then, one has to extract, among
these eigenfunctions (by the direct action of $Q^-$), those which
are simultaneously the normalizable zero modes of $Q^-$. If this
plan can be realized, we will reduce the spectral problem for the
Hamiltonian $H_1$ (which is not amenable to conventional
separation of variables) to a couple of one-dimensional spectral
problems.

Keeping in mind \cite{new} that sometimes the preliminary
similarity transformations are helpful for separation of
variables, we will transform the operator $Z$ using a function
$\exp\{\varphi(\vec{x})\}$ that will be determined later on: \ba Y
&\equiv & e^{-\varphi(\vec{x})} Z e^{\varphi(\vec{x})}=(\beta
-\alpha )\partial_1^2-(\alpha + \beta a^2)\partial_2^2+ 2(\beta
-\alpha)(\partial_1\varphi)\partial_1-\nonumber\\&&-2(\alpha +
\beta a^2)(\partial_2\varphi)\partial_2 -\beta
C_k\partial_k+(\beta -\alpha)\biggl((\partial_1^2\varphi )
+(\partial_1\varphi)^2\biggr)-\nonumber\\&&-(\alpha + \beta a^2)
\biggl((\partial_2^2\varphi )+(\partial_2\varphi)^2\biggr)-\beta
C_k\partial_k\varphi +\beta (B-\partial_kC_k)+bV_1.
  \label{ZY}
\ea

To exclude the first order derivatives in $Y,$ we have to impose
two conditions: \ba 2(\beta - \alpha )(\partial_1\varphi )=\beta
C_1 \qquad -2(\alpha +\beta a^2)(\partial_2\varphi )=\beta C_2
\label{33} \ea which are satisfied if
\begin{equation}
(\alpha + \beta a^2)\partial_2C_1 = -(\beta -\alpha)\partial_1C_2.
\nonumber
\end{equation}
Comparing with (\ref{1.3}), it is clear that the two constants
introduced in (\ref{linear}) are related by $2\alpha =\beta
(1-a^2).$

Substituting now $\partial_1\varphi,
\partial_2\varphi$ from (\ref{33}) into
(\ref{ZY}), we get the expression: \ba Y=\frac{\beta
(1+a^2)}{2}\biggr(\partial_1^2-\partial_2^2+\frac{\partial_kC_k}{1+a^2}-
\frac{C_1^2-C_2^2}{(1+a^2)^2}\biggl)+\beta
\biggl(B-\partial_kC_k+\frac{1-a^2}{2}V_1\biggr). \label{35} \ea
Taking into account that from (\ref{1.2})
$\partial_kC_k=(1-a^2)(V_1-V_2)/2$ and also (\ref{8}), one
obtains: \ba B-\partial_kC_k+\frac{1-a^2}{2}V_1 =
B+\frac{1-a^2}{2}V_2=-\frac{1}{a^2}C_+C_-+
\frac{1+a^2}{2}(F_2-F_1). \nonumber \ea Now using (\ref{CCv}), we
can write (\ref{35}) in the form: \ba Y= \frac{\beta
(1+a^2)}{2}\Biggl(\partial_1^2-\partial_2^2 -\frac{1-a^2}
{a(1+a^2)}\biggl(C^{\prime}_++
\frac{C_+^2}{a(1+a^2)}-C^{\prime}_-+\frac{C_-^2}{a(1+a^2)}\biggr)+F_2-
F_1\Biggr). \label{37} \ea

Using the expressions (\ref{C}) - (\ref{CCv}) for $C_i,$ the
function $\varphi (\vec x)$ can be written explicitly in terms of
$C_{\pm}(x_{\pm}):$
\begin{equation}\label{phi}
\varphi (\vec x) = \frac{-1}{a(1+a^2)}\biggl(\int C_+(x_+)dx_+-
\int C_-(x_-)dx_- \biggr).
\end{equation}

Though for generic functions $C_{\pm}(x_{\pm})$ variables in
(\ref{37}) can not be separated, for some of ans\"atze considered
before this is just possible due to a specific choice of
$C_{\pm}.$ In particular, the second ansatz of Subsection 3.1 with
$\eta\neq 0,$ $C_-=0$ and $C_+$ satisfying (\ref{ricatti}),
eliminates all obstacles to separate variables for (\ref{37}) in
terms of $x_1,x_2.$ The same happens for the ansatz of the
Subsection 3.2, if one of the constants $\sigma_+, \delta_+$
vanishes. In this case the functions $C_{\pm}(x_{\pm})$ satisfy
also the Ricatti equation (\ref{a2}), and the potential $V_2$ in
(\ref{299}) allows separation of variables.

In both of these models, the similarity transformation above
separates variables, and the eigenfunctions of the operator $Y$
(and therefore, of $Z$) can be built from the eigenfunctions of
the corresponding pair of one-dimensional problems.

Now, we will briefly consider the particular case of the model
with partner potentials (\ref{29}) - (\ref{299}) for $\sigma_+=0.$
Then, the physical system is described as
\begin{equation}
H_2=h_1(x_1)+h_2(x_2)\qquad h_1=-\partial_1^2+F_1(x_1) \qquad
h_2=-\partial_2^2+F_2(x_2) \nonumber
\end{equation}
\begin{equation}
h_1\psi_1(x_1)=\epsilon_1\psi_1(x_1) \qquad
h_2\psi_2(x_2)=\epsilon_2\psi_2(x_2) \nonumber
\end{equation}
where the two one-dimensional Hamiltonians $h_1$ and $h_2$ are, up
to additive constants $c_1,c_2,$ Morse potentials with well known
bound states, if $\sigma_-\delta_->0.$ For simplicity, we will
consider $\sigma_-=\delta_-\equiv \sigma $ and $d<0.$ Substituting
(\ref{Fff}) into (\ref{CCv}) and (\ref{8}) one obtains the
expression for the supercharge $Q^+:$ \ba
Q^+=-h_1+a^2h_2+C_i\partial_i
+\frac{\sqrt{-da(1+a^2)}}{a^2}C_-+\frac{(1-a^2)d}{a}. \nonumber
\ea The action of this operator on the wave functions
$\Psi(\vec{x})=\psi_1(x_1)\psi_2(x_2)$ of $H_2$ will give the wave
functions (if normalizable) of the partner $H_1:$ \ba
 Q^+\Psi (\vec x)&=&\biggl(a^2\epsilon_2-\epsilon_1+\frac{(1-a^2)d}{a}\biggr)\Psi (\vec x)
- \frac{\sqrt{-da(1+a^2)}}{a}(\partial_1 - a \partial_2)\Psi (\vec
x) +\nonumber\\
  &+&\frac{C_-(x_-)}{a}\biggl( \partial_1 + a\partial_2
  -\frac{\sqrt{-da(1+a^2)}}{a} \biggr)\Psi (\vec x). \label{x2}
\ea The one-dimensional wave functions $\psi_1(x_1)$ and $
\psi_2(x_2)$ are known explicitly (see, for example \cite{new}) in
terms of hypergeometric functions. One can straightforwardly check
that the singularity of $C_-(x_-)$ at $x_-=0$ in the last term of
(\ref{x2}) can not be compensated. Therefore, the bound states
$\Psi (\vec x)$ of $H_2$ have no normalizable partner states in
the spectrum of $H_1$. According to the discussion carried out in
the the first part of this Section, the additional bound states of
$H_1$ could be found among normalizable zero modes of $Q^-.$ In
the considered model, the operator (\ref{37}) takes a simple form
(up to a constant factor and a constant additive term), which
allows the separation of variables:
\begin{equation}
Y\sim -h_1(x_1)+h_2(x_2)+\frac{2(1-a^2)}{a(1+a^2)}. \nonumber
\end{equation}
Then, the eigenfunctions $\Psi_Z(\vec x)$ of the operator $Z$,
defined in (\ref{linear}), are:
\begin{equation}
\Psi_Z=e^{\varphi(\vec x)} \Psi_Y(\vec x) =e^{\varphi(\vec x
)}\cdot\psi_1(x_1)\psi_2(x_2) \nonumber
\end{equation}
where $\psi_1(x_1), \psi_2(x_2)$ are arbitrary eigenfunctions of
$h_1, h_2,$ respectively, and the similarity transformation is
performed by
\begin{equation}\label{x5}
e^{\varphi(\vec x )}\sim\frac{\exp(-\gamma x_+)}{\sinh(\gamma
x_-)}.
\end{equation}
The singularity of (\ref{x5}) at $x_-= 0$ is responsible of the
non-normalizability of all eigenfunctions $\Psi_Z(\vec x)$,
including possible zero modes of $Q^-$. Therefore, the Hamiltonian
$H_1$ has no bound states at all, despite the presence of two
Morse potentials in (\ref{29}).

\section{Two-dimensional harmonic oscillator}
\vspace*{0.1cm} \hspace*{3ex} In the previous successful attempts
\cite{david} - \cite{exact} to solve the two-dimensional
intertwining relations of second order (\ref{intertw}), the system
of nonlinear differential equations (\ref{1}) - (\ref{3}) was
solved step by step, and only at the very end some specific
expressions for potentials were obtained. Up to now, it has been
impossible to solve these equations straightforwardly, starting
from the fixed form of one of the partner potentials, apart from
some very simple examples with separation of variables in both
$V_1$ and $V_2$. The soluition of the problem is not so evident in
the case of supercharges with deformed hyperbolic metric. In this
Section, we will take $V_2(\vec x)$, as an isotropic harmonic
oscillator: \ba
  V_2(\vec x)=\Omega (x_1^2 + x_2^2) +v_0=\omega
  (x_+^2+x_-^2)+2\mu x_+x_-+v_0 \label{osc}\\
  x_{\pm}=x_1\pm
  ax_2 \qquad \omega = \frac{\Omega (1+a^2)}{4a^2}\qquad \mu = -\frac{\Omega
  (1-a^2)}{4a^2}. \nonumber
\ea One can check that non-isotropic oscillator does not produce
any solution of the intertwining relations. Substituting this
potential in (\ref{consist}), we obtain
\begin{equation}
  \partial_+^2[C_+^2 +\sqrt{-b}C_+^{\prime}] =
  \partial_-^2[C_-^2 -\sqrt{-b}C_-^{\prime}]\equiv 2A\qquad
  A=const  \nonumber
\end{equation}
where $A$ is an arbitrary constant, and $b$ was defined in
(\ref{K}). The variables are separated, and the functions
$C_{\pm}(x_{\pm})$ satisfy the Ricatti equations: \ba
C_+^2(x_+)+\sqrt{-b}C^{\prime}_+(x_+)= A^2x_+^2+A_+x_++a_+
\label{osc1}\\
C_-^2(x_-)-\sqrt{-b}C^{\prime}_-(x_-)= A^2x_-^2+A_-x_-+a_-.
\label{osc2} \ea Then, equations (\ref{5}) - (\ref{6}) give the
expression for the function $B(\vec x)$ in terms of $C_+,C_- :$
\ba
  B(\vec x) &=& -\frac{1}{a^2}C_+C_- - \frac{\mu (1+a^2)+\omega
  (1-a^2)}{2}(x_+^2+x_-^2)+ \nonumber\\&+&\frac{A-a^2(\mu (1-a^2)+\omega
  (1+a^2))}{a^2}x_+x_-+\rho_+x_++\rho_-x_-+const. \nonumber
\ea

After rather cumbersome but straightforward calculations, one can
check that the last equation to be solved (\ref{3}), is fulfilled
only if, in addition to (\ref{osc1}) - (\ref{osc2}), one of the
functions $C_+,C_-$ is linear, for example $C_-(x_-)=\sigma x_-.$
Then, we finally obtain that
\begin{equation}\label{final}
  V_1(\vec x)= V_2(\vec x) -\frac{2}{a}C_+^{\prime}(x_+)+ const.
\end{equation}
The Schr\"odinger equation with such potential $V_1$ {\it does not
allow} separation of variables in terms of $x_{\pm}$ due to the
presence of a mixed term in the Laplacian, but it obviously {\it
allows} separation in the variables
$z_+=x_+=x_1+ax_2,\,\,z_-=ax_1-x_2,$ due to the absence of mixed
term. Therefore, no non-trivial isospectral superpartners of the
isotropic harmonic oscillator can be built in the framework of
second order intertwining relations for superchrages with constant
metric.

\section{Conclusions}
\vspace*{0.1cm} \hspace*{3ex} The investigation of intertwining
relations (\ref{intertw}) with deformed hyperbolic metric
$g_{ik}=diag(1,-a^2),\quad a\neq 0,\pm 1$  in the supercharge
operator (\ref{ourq}), that has been carried out in this paper,
completes the study of second order intertwining relations with
{\it constant matrix $g_{ik}$} in two-dimensional Quantum
Mechanics, which was started in \cite{david} - \cite{exact}. In
the present case also some particular classes of solutions for the
partner potentials were found (see Section 3). Among them, there
are pairs in which one of the partners allows separation of
variables, but the second one does not. A specific procedure of
SUSY-separation of variables is proposed for this case (see
Section 4). In a particular model, this new algorithm led to the
conclusion that the system with the attractive potential
(\ref{29}) for $\sigma_+=0,\,\sigma_-=\delta_-$ does not allow any
bound states. Although the nonlinear equation (\ref{consist}) is
not amenable, as a rule, to solution with given potential $V_2$
(e.g. for the Coulomb potential), sometimes this is possible. Thus
in the case of one partner being harmonic oscillator, it is shown
in Section 5 that the second one also allows separation of
variables. All models considered in the paper are completely
integrable, i.e., nontrivial symmetry operators of fourth order in
momenta $(R_1=Q^+Q^-;\, R_2=Q^-Q^+)$ exist for the deformed
hyperbolic metric $g_{ik}$ also.

A few additional ans\"atze could be considered in the same manner.
In particular, one can check that the case $C_-(x_-)=Const\neq 0$
(the natural generalization of the Subsection 3.1) leads to a
particular solution, obtained above within the ansatz in the
Subsection 3.2. The partner of the constant potential $V_2=Const$
also allows separation of variables, similar to the case
considered in Section 5. In this paper we were only interested in
real potentials. Some models with complex potentials (see
\cite{pseudo}, \cite{exact}) can be considered if for example we
allow purely imaginary $a\neq\pm i$ or arbitrary values of $\nu ,
\tilde\nu $ in (\ref{Fnu}).

\section{Acknowledgements}

This work has been partially supported by Spanish Ministerio de
Educaci\'on y Ciencia under Projects SAB2004-0143 (sabbatical
grant of MVI), MTM2005-09183 and Junta de Castilla y Le\'on
(Excellence Project VA013C05). The research of MVI was also
supported by the Russian grants RFFI 06-01-00186-a, RNP 2.1.1.1112
and RSS-5538.2006.2. 

\end{document}